\documentclass[prc,twocolumn,nofootinbib]{revtex4-2} 
\usepackage{color}

\usepackage{amsmath}
\usepackage{amssymb}
\usepackage{graphicx}
\usepackage{bm}
\usepackage{hyperref}
\usepackage{url}
\hypersetup{colorlinks = true, allcolors = blue}

\usepackage[normalem]{ulem}

\begin{document}

\title{Non-empirical shape dynamics of heavy nuclei with multi-task deep learning}

\author{N. Hizawa}
\author{K. Hagino}
\affiliation{
Department of Physics, Kyoto University, Kyoto 606-8502,  Japan}
\begin{abstract}
A microscopic description of nuclear fission represents one of the most challenging problems in nuclear theory.
While phenomenological coordinates, such as multipole moments, have often been employed 
to describe fission, it is not obvious whether these parameters fully reflect the shape dynamics of interest. 
We here propose a novel method to extract collective coordinates, which 
are free from phenomenology, based on multi-task deep learning in conjunction with 
a density functional theory (DFT).
To this end, we first introduce randomly generated external fields to a Skyrme-EDF 
and 
construct a set of nuclear number densities and binding energies 
for deformed states of ${}^{236}$U around the ground state. 
By training a neural network on such dataset with
a combination of an autoencoder and supervised learning, 
we successfully identify a two-dimensional latent variables that accurately reproduce both the energies and the 
densities of the original Skyrme-EDF calculations, within a mean absolute error
of 113 keV for the energies.
In contrast, when multipole moments are used 
as latent variables for training in constructing the decoders, 
we find that the training data for the binding energies are reproduced only within 2 MeV. 
This implies that 
conventional multipole moments do not provide fully adequate variables for a shape dynamics of 
heavy nuclei. 
\end{abstract}

\maketitle

\section{Introduction}
What are the most consistent collective coordinates to describe a shape dynamics of heavy nuclei, such as 
nuclear fission? And how many collective coordinates do we need to achieve an essential description of 
such shape dynamics? These are the questions which we discuss in this paper. 

Nuclear fission is in fact one of the most fundamental and yet challenging problems in nuclear 
theory. In most of calculations carried out so far, except for 
the ones based on the time-dependent density functional theory (TDDFT) \cite{Bulgac2016}, 
one usually assumes a few collective coordinates for fission 
and constructs a (multi-dimensional) potential energy surface \cite{Schunck2022}. For a spontaneous fission, 
one in addition computes the moment of inertia and estimates the decay half-life based on the WKB 
approximation \cite{Brack72,Schunck2016,Sadhukhan2022,Flynn2022,Staszczak2009,Washiyama2021}. 
On the other hand, for induced fission, one computes transport coefficients and solves the dynamics with e.g., the Langevin 
approach \cite{Aritomo2013,Ishizuka2017}, the random walk approach \cite{Randrup2011,Randrup2011b}, 
and the time-dependent generator coordinate method (TDGCM) \cite{Goutte2005,Regnier2019,Zhao2022}. 

The most serious problem in these approaches is that 
one has to assume a priori relevant collective degrees of freedom for fission based on a phenomenological 
consideration, despite that collective degrees of freedom for large amplitude motions, such as fission,  
are highly nontrivial. Therefore, it is not obvious whether chosen collective coordinates are optimum 
for the dynamics of interest. 
In principle, 
the self-consistent collective coordinate method (SCC) \cite{SCC1, SCC2, ASCC} can derive a collective path 
without resorting to a phenomenological assumption. 
Even though 
several approximate calculations based on SCC have been carried out \cite{HN08, WN17}, 
it is still computationally quite demanding. 
Evidently, a more efficient method to determine collective coordinates is urged. 

We believe that a machine learning technique will provide a useful means in this respect. 
As it continues to evolve, 
machine learning is bringing a pivotal shift in nuclear theory.
Currently, regression analysis with supervised learning is a common practice 
in nuclear theory \cite{LR20, SJ21, WP21, WG21, WL22, MS22, LM22, ML22, WR22, MT23, ZL23, NB23, ZY23, BD22}. For example, 
it has been employed to construct a new mass table \cite{Wu2020}. 
It is important to notice that 
machine learning extends beyond supervised learning, 
encompassing other techniques such as unsupervised learning \cite{unsupervise_review}, reinforcement learning \cite{reinforce_review}, and self-supervised learning \cite{self_super_review}. 
These have successfully been applied not only in physics but also in several other 
disciplines as well.

Of particular interest within unsupervised learning is a method of an autoencoder \cite{recnet_AE}. 
In this method, 
input data are mapped onto low-dimensional latent variables, with which the input data are subsequently 
reconstructed back.
The underlying idea of this method is based on the manifold hypothesis \cite{FM16}, 
that is, data of interest occupy only limited locations in a quite high-dimensional space.
This technique has been applied in various contexts, including a data characteristic analysis and a 
computation cost reduction.

In nuclear physics, Ref. \cite{VS22} recently employed autoencoders to acquire low-dimensional 
representations of quadrupole and octupole deformed Hartree-Fock-Bogoliubov (HFB) wave functions, which 
were to be used in Generator Coordinate Method (GCM) calculations. Notice that the manifold hypothesis 
may hold rather trivially in the study of Ref. \cite{VS22}, as autoencoders were applied only to limited datasets. 
Nevertheless, the basic idea of Ref. \cite{VS22} remains intriguing. 
In fact, the concept of collective coordinates in nuclear physics is practically 
equivalent to the manifold hypothesis.
That is, both of them state that 
low-energy dynamics of a nucleus can be approximated within a restricted region 
in the infinite-dimensional manifold which fully specifies degrees of freedom of a system, such as the full 
Hilbert space. 
It would thus be reasonable to anticipate that machine learning methods, including unsupervised learning, work effectively 
in describing nuclear collective motions.

In this paper, we shall apply multi-task learning (MTL) \cite{MTL_review, WL22, ZY23} to discuss adiabatic shape dynamics of $^{236}$U near the ground state.
To this end, in this work we employ 
the Density Functional Theory (DFT) \cite{Kohn-Sham, Hohenberg-Kohn} and 
combine an autoencoder and supervised learning. 
A remarkable advantage of DFT is that the only dynamical variable is in principle a particle number density.
This is in contrast to other many-body theories with many-body wave functions,  
which are significantly more difficult to handle even numerically.
By employing multi-task learning, we shall 
extract a common feature representation of the energy and the density of $^{236}$U,
that is, common latent variables which yield both the energy and the density.  
This representation is expected to contain a large amount of information on the adiabatic dynamics.

The paper is organized as follows. 
In Sec. II, we will detail our procedure for MTL. To construct a dataset, we will 
introduce random potentials to the system. 
In Sec. III, 
we will discuss criteria for evaluations of our MTL calculations. 
We will apply these criteria to discuss the applicability of the conventional 
collective coordinates for fission, 
by comparing the results with the criteria. 
In Sec. IV, 
we will carry out MTL and discuss latent variables for shape dynamics. 
We will then summarize the paper in Sec. V and discuss future perspectives.  

\section{Formulation}

\subsection{Skyrme EDF with random external potentials}

Our purpose in this paper is to analyze a low-energy deformation dynamics of $^{236}$U with the Density Functional Theory (DFT) \cite{Hohenberg-Kohn, Kohn-Sham} 
and multi-task learning.
The DFT evaluates a ground state dynamics, i.e., the adiabatic dynamics, under arbitrary external fields. 
An elemental degree of freedom in the DFT is a particle number density $\rho$, 
with which any observable, including the ground state energy $E$, can be evaluated.
Namely, in principle, there is an energy density functional (EDF) $E[\rho]$, that solely depends on the particle number density, 
to describe the adiabatic dynamics.
However, in nuclear physics, almost all calculations employ a Kohn-Sham type energy density functional (KS-EDF), 
which depends not only on the particle number density but also on other densities, such as the kinetic energy, the spin-orbit, and 
the pairing densities \cite{BH03}. 
Among them, the Skyrme EDF is one of the most famous and commonly used KS-EDFs \cite{BH03,VB72,Ring_Schuck}, and we shall also adopt it 
in this paper.
For simplicity, we impose the axial and the time reversal symmetries on the system.
The actual calculations are performed with the computer code {\tt HFBTHO} (v3.00) \cite{HFBTHOv3.00}. This is an open source DFT solver, 
in which Kohn-Sham wave functions are expanded on an axial symmetric harmonic oscillator basis,
\begin{gather}
    \varphi_{n_z, n_r,\Lambda}(\bm{r};b_z, b_{\perp}) 
     =
    \varphi_{n_z}(\xi;b_z)
    \varphi^{|\Lambda|}_{n_r;b_{\perp}}(\eta)
    \frac{e^{i\Lambda\phi}}{\sqrt{2\pi}} \label{eq:basis}\\
    \varphi_{n_z}(\xi;b_z)
     = 
     \frac{1}{\sqrt{b_z2^{n_z}n_z!\sqrt{\pi}}}H_{n_z}(\xi)e^{\xi^2/2} \\
     \varphi^{|\Lambda|}_{n_r}(\eta;b_{\perp})
     = 
     \sqrt{\frac{2n_r!}{b_{\perp}^2(n_r+|\Lambda|)!}}\
     \eta^{|\Lambda|/2}L_{n_r}^{|\Lambda|}(\eta)e^{-\eta/2}  \\
     \xi = \frac{z}{b_z}, \quad 
     \eta = \left(\frac{r_\rho}{b_{\perp}}\right)^2
\end{gather}
where $\bm{r} = (r_\rho, z, \phi)$ is a three-dimensional cylindrical coordinate, and $b_{\perp}$ and $b_z$ are the 
oscillator length parameters.
$H_{n_z}$ and $L_{n_r}^{|\Lambda|}$ are the Hermite polynomials and the associated Laguerre polynomials, respectively.
Note that the octupole deformation is included in our analysis, as 
we do not impose the reflection symmetry along the $z$ axis. 

We adopt the SLy4 parameter set \cite{SLy4} and a surface type paring interaction \cite{BH03}
to evaluate the Kohn-Sham DFT within the Hartree-Fock-Bogoliubov (HFB) scheme \cite{Ring_Schuck}.
The strengths of the pairing interaction are determined 
so that the average pairing gaps of protons and neutrons agree with the empirical value $12/\sqrt{A}$ MeV 
with the mass number $A=236$. The resultant values are 
$-650.98$ MeV/$\mathrm{fm}^3$ for protons and $-526.53$ MeV/$\mathrm{fm}^3$ for neutrons.
We also optimize $b_z$ and $b_{\perp}$ to minimize the ground state energy of $^{236}$U, with the maximum 
number of oscillator shells $N_{\mathrm{tot}}$ of 26, which is large enough to describe the deformation dynamics 
around the ground state of $^{236}$U.
We obtain $b_z = 2.08$ fm and $b_{\perp} = 1.92$ fm, 
with which the ground state energy is found to be $-1780.87$ MeV.

To construct a set of the densities and the energies, 
$(\rho, E)$, one must solve the Kohn-Sham equation with various external fields $v$.
Even though the density constraint method \cite{Umar2006} directly provides the energy 
for a desired density, a numerical cost in this method may increase when a numerical accuracy 
is required. 
We shall therefore adopt a random potential approach \cite{MS17} in this paper.

Notice that it is computationally impossible to deal with fully arbitrary densities. Even if that was possible, 
such densities might contain enormous amount of information that is not of our interest, 
especially those which correspond to high energy configurations. 
Therefore, it is desirable to sample densities according to a probability distribution $p[\rho]$ 
such that a large portion of the sampled densities are relevant to our purpose. 
However, what one actually does in the KS-EDF is to sample external fields from some probability distribution $q[v]$ that is 
normalized as, 
\begin{equation}
    \int \mathcal{D}v\,q[v] := 1,
\end{equation}
with the functional integration of $v$.

In principle, one can map the probability distribution $q[v]$ onto the probability distribution 
 $p[\rho]$ in the following way. 
Since there is a bijection $\rho[v]$ and $ v[\rho]$ in the DFT, the conditional probability is formally denoted 
by $P(\rho|v) = \delta[\rho - \rho[v]]$, where the delta function is normalized as
\begin{equation}
    \int\mathcal{D}\rho \,\delta[\rho - \rho[v]] := 1, \quad \int \mathcal{D}\rho\,p[\rho] := 1
\end{equation}
Therefore, sampling $v$ from $q[v]$ yields the probability distribution,
\begin{equation}
    p[\rho]
    =
    \int \mathcal{D}v\, \delta[\rho - \rho[v]] q[v]
 =
    \left[
        \left|
        \mathrm{Det}\left(\frac{\delta v}{\delta \rho}\right)
        \right|
        q[v]
    \right]_{v = v[\rho]}, 
\end{equation}
where $\mathrm{Det}$ is a functional determinant.

Since the relation between the densities and the external fields, that is, $\rho[v]$ and $v[\rho]$, is nontrivial, 
it is a highly complicated problem how one can choose external potentials appropriately. 
Although the strategy of selecting an appropriate $q[v]$ is unknown, it is meaningful 
to sample $v$ highly randomly.
It is important to notice here that in general 
the result of a machine learning with a dataset sampled from $q[v]$ may not work well 
with another dataset generated from a different probability distribution $q'[v]$.
This is recognized as a serious issue, referred to as the problem of the domain shift, 
in the field of machine learning \cite{domain_adaptation_survey1, domain_adaptation_survey2}.
In particular, in the field of DFT, it has been shown that there can be a case where results trained with a {\it simple} $q[v]$ 
have little predictive ability when it is applied to highly random data with $q'[v]$ \cite{RS19, HH23}.
On the other hand, results with a random data with $q'[v]$ are expected to have a much better predictive ability, 
and thus we employ random potentials in this study. 
To this end, we introduce random potentials 
based on the axial harmonic oscillator basis given by Eq. (\ref{eq:basis}): 
\begin{equation}
    v^{(i)}(r, z)
    =
        \sum'
    v^{(i)}_{n_z, n_r, |\Lambda|}
    \varphi_{n_z}(\xi;b^{(i)}_z)
    \varphi^{|\Lambda|}_{n_r}(\eta;b^{(i)}_{\perp}), 
    \label{eq:v}
\end{equation}
where $\sum'$ indicates that the sum is restricted with a condition 
$n_z/q + 2n_r + |\Lambda|\leq N_{\mathrm{shell}}$, with $q=(b_z/b_{\perp})^2=1.1736$. 
Here, 
$i$ denotes a index of data, for which  
$v^{(i)}_{n_z, n_r, |\Lambda|}$, $b^{(i)}_z$, and $b^{(i)}_{\perp}$ are randomly generated.
Note that, in addition to the weight factors $v_{n_z, n_r, |\Lambda|}$, we also randomly sample the length parameters, $b_z$ and $b_{\perp}$, 
even though the basis functions can express arbitrary function with fixed values of $b_z$ and $b_{\perp}$.
This is to avoid the neural network to learn the specific scales $(b_z, b_{\perp})$ when adopting a finite shell number $N_{\mathrm{shell}}$.
We apply the same external fields both to protons and neutrons. 

The random potentials (\ref{eq:v}) are not symmetric along the $z$ axis, and the center of mass position of the system 
can move around.
To fix it, we add the quadratic constraint on the center of mass position to the KS-EDF during the HFB 
iterations\footnote{Although not done in this paper, an alternative way to fix the center of mass position is to introduce the Lagrange multiplier. This operation corresponds to restricting the domain of the random potentials.}:
\begin{equation}
    E_{\mathrm{CoM}}[\rho]
    =
    \frac{C}{2}\left(
    \int d^3 r\,z\rho(\bm{r})
    -
    z_0 
    \right)^2, 
\end{equation}
with $C=1.0$ MeV fm$^{-2}$ and $z_0=0$. 
When one needs a dataset in which the center of mass position is not fixed, 
one can simply add data by shifting the densities. Because of the translational invariance in the nuclear Hamiltonian, 
the total binding energy will remain the same. This approach reduces the cost of creating a new dataset and maintaining it.

The basic strategy to determine the random potentials 
is to generate a large number of deformed states near the ground state, 
with excitation energies below about $15$ MeV, 
which is larger than the first barrier for fission.
To this end, 
we scale the optimized oscillator lengths as $b^{(i)}_{z, \perp} = (1 + t^{(i)}_{z, \perp})b_{z, \perp}$ 
with uniform random numbers of $t^{(i)}_{z, \perp}$ within $[0, 0.5]$.
On the other hand, we individually generate $v^{(i)}_{n_z, n_r, |\Lambda|}$ for each set of $(n_r, n_z, |\Lambda|)$ 
using uniform random numbers within $[-v_0, v_0]$.
Note that a single scale cut off parameter $v_0$ leads to a 
significant bias in the distribution of the calculated binding energies.
To avoid this problem, 
we adopt a multiple values of $v_0$, that is, 0.6 MeV fm$^{3/2}$, 1.2 MeV fm$^{3/2}$, 1.8 MeV fm$^{3/2}$, and 2.4 MeV fm$^{3/2}$.  
Moreover, 
a certain set of data should be removed so that the resultant data do not have a large bias 
in the binding energies.
For this purpose, we randomly remove excessive amount of data at each bin in the histogram 
when the number of data exceeds 2250. The total number of bins is taken to be 100. 
Furthermore, we also discard datasets outside of $-1780.87 \ \mathrm{MeV}\leq E \leq -1765.00 \ \mathrm{MeV}$.
In addition, as we are interested in the shape dynamics only around the ground state in this paper, 
we reject the data with large deformation, with a cut-off of [3.0 b, 16.0 b] for an expectation value of the mass 
quadruple moment operator, 
\begin{equation}
    \hat{Q}_{\lambda k} = r^{\lambda}Y_{\lambda k}(\hat{\theta}, \hat{\phi})
\end{equation}
with $\lambda=2$ and $k=0$, where $Y_{\lambda k}$ is the spherical harmonics with the polar coordinates $\bm{r} = (r, \theta, \phi)$.

\begin{figure}[tbp]
    \begin{center}
    \includegraphics[width=86mm]{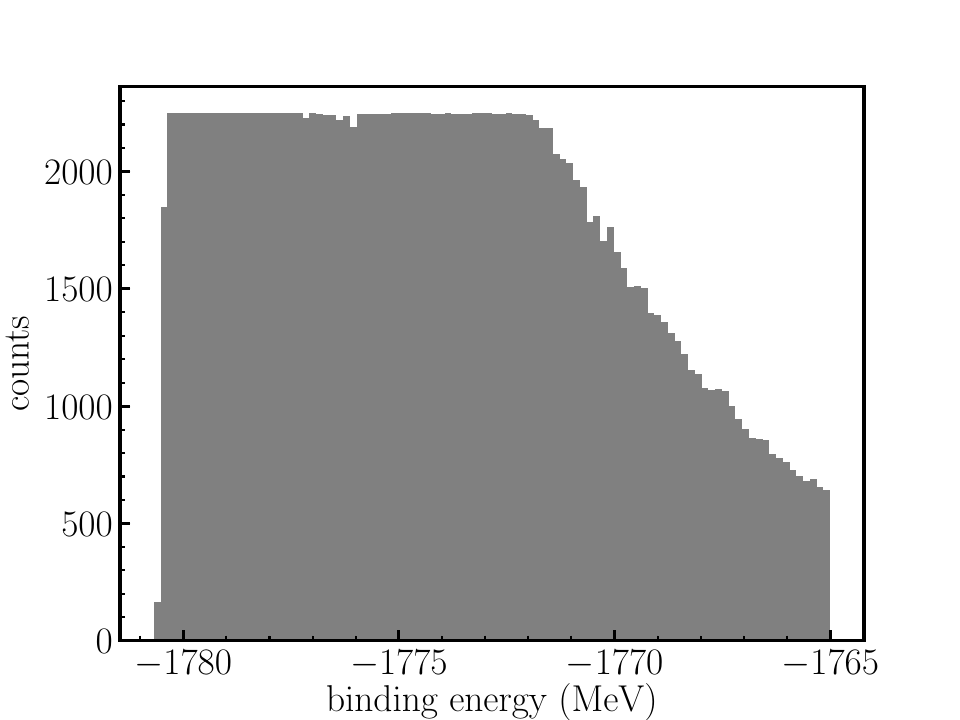}
    \caption{ \label{fig:hist_energy}
    A histogram of the binding energies for the dataset generated in this work.
    Since we randomly remove excessive amount of data, the distributional bias is largely eliminated.
    The ground state energy is $-1780.87$ MeV, and one can see that 
    the data are almost uniformly distributed 
    below the excitation energy of 8 MeV, i.e., the energy of $-1772$ 
    MeV. 
    }
    \end{center}
\end{figure}

\begin{figure}[thbp]
    \begin{center}
    \includegraphics[width=86mm]{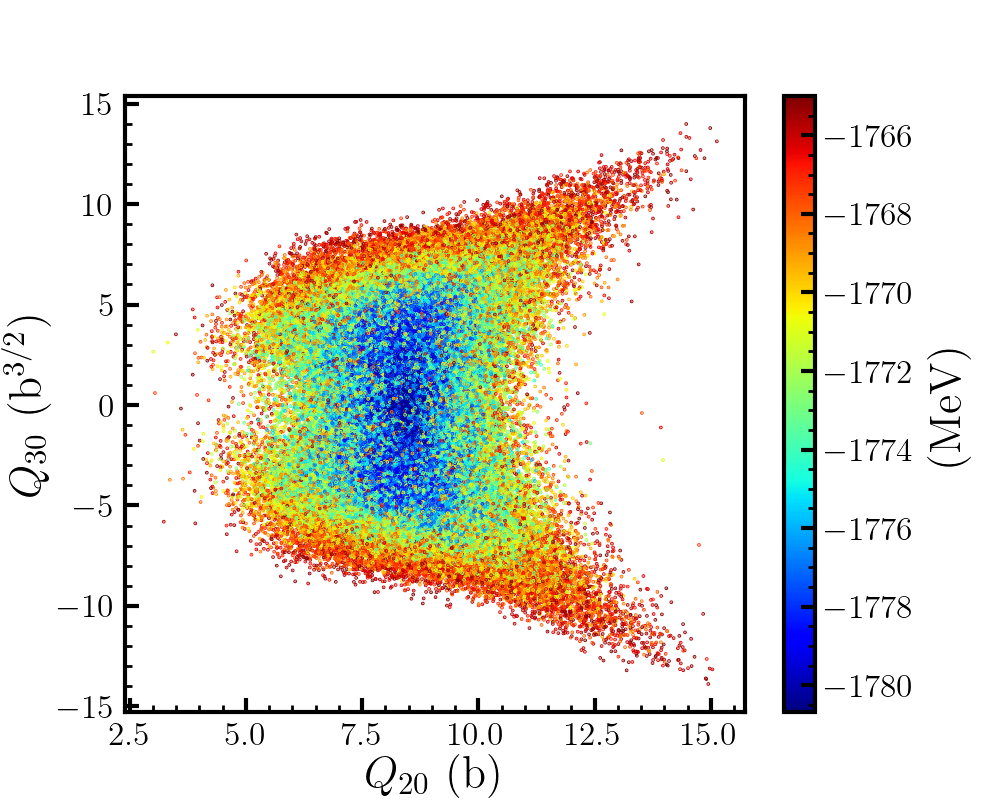}
    \caption{\label{fig:scatter_data}
    A scatter plot for the quadruple and the octupole moments for the dataset.
    The color of each points shows the corresponding biding energy.
    The image appears symmetric along $Q_{30}=0$, reflecting the fact that the results of the Skyrme-EDF 
    are invariant under the parity transformation with respect to the $z$ axis.
    }
    \end{center}
\end{figure}

With this strategy, we generate 298982 random potentials, out of which 181134 potentials 
remain for the machine learning. 
The composition of each $v_0$ is 20 \% for $v_0=$ 0.6 MeV fm$^{3/2}$, 23 \% for $v_0=$ 1.2 MeV fm$^{3/2}$, 
49 \% for $v_0=$ 1.8 MeV fm$^{3/2}$, and 8 \% for $v_0=$ 2.4 MeV fm$^{3/2}$.
Figures \ref{fig:hist_energy} and \ref{fig:scatter_data} show the histogram of the binding energies and 
the energy surface for the remaining data, respectively. 
For the latter, we plot the energy for each density in the two-dimensional plane for the quadrupole and the 
octupole moments corresponding to the density. 
Note that the binding energy is calculated excluding the energy for the external fields. 
To utilize the densities for deep learning, we discretize them with the mesh size 0.4 fm for 
both the $z$ and $r$ axes, which is approximately half of values commonly used in Skyrme DFT calculations with a 3D lattice representation.
The number of mesh points are 48 points (for the $r$ axis), and 128 points (for the $z$ axis).
The nucleon density is then regarded as an monotonic image of $48\times 128$ pixels.
Notice that it is desirable to have the number of pixels with a power of 2, 
since layers that double or halve the number of vertical and horizontal pixels are often used
in the field of image recognition. 
Our choice of mesh points approximately satisfies this condition.

\subsection{Deep learning}

\subsubsection{multi-task learning}

Based on the manifold hypothesis and the assumption of collective coordinates in nuclear theory, we expect that 
our dataset constructed in the previous subsection can be well characterized by a small number of parameters.
To extract such latent variables of a dataset,   
the autoencoder is a common approach \cite{recnet_AE}. 
We apply this to extract latent variables for our dataset. 
Namely, an input density is encoded to latent variables and then 
a decoder attempts to reproduce the original density with the latent variables as inputs. 
This is schematically denoted as
\begin{equation}
    \rho\to\bm{z} = \mathcal{E}[\rho]\to \rho = \mathcal{D}(\bm{z}),
    \label{eq:learning-rho}
\end{equation}
where $\bm{z} = \{z_1, z_2, \cdots, z_{d_{z}}\}$ are the extracted latent variables with dimension $d_z$
and  $\mathcal{E}$ and $\mathcal{D}$ are an encoder and a decoder, respectively. 
The number of extracted latent variables may become too large to understand if a high accuracy is imposed onto the 
encoder-decoder system. In practice one would thus need a compromise for the accuracy. 
In addition, a large error may be generated when 
reconstructed densities are applied to predict energies with the 
orbital-free EDF (OF-EDF). This problem is known as 
the density driven error (DDE) \cite{BV17}, 
and in this case the latent variables may not have enough information on the energies.

On the other hand, one can directly apply the encoder-decoder structure to learn the OF-EDF itself.
Namely, one can construct a decoder which reproduces the energy, rather than the original density, as 
\begin{equation}
    \rho\to\bm{z} = \mathcal{E}[\rho]\to E = \mathcal{D}(\bm{z}). 
\end{equation}
However, in this supervised learning approach, there is always a risk of leading to 
meaningless variables, for example, 
when all of the latent variables are regarded as a transformation of the energy.

\begin{figure}[tb]
    \begin{center}
    \includegraphics[width=86mm]{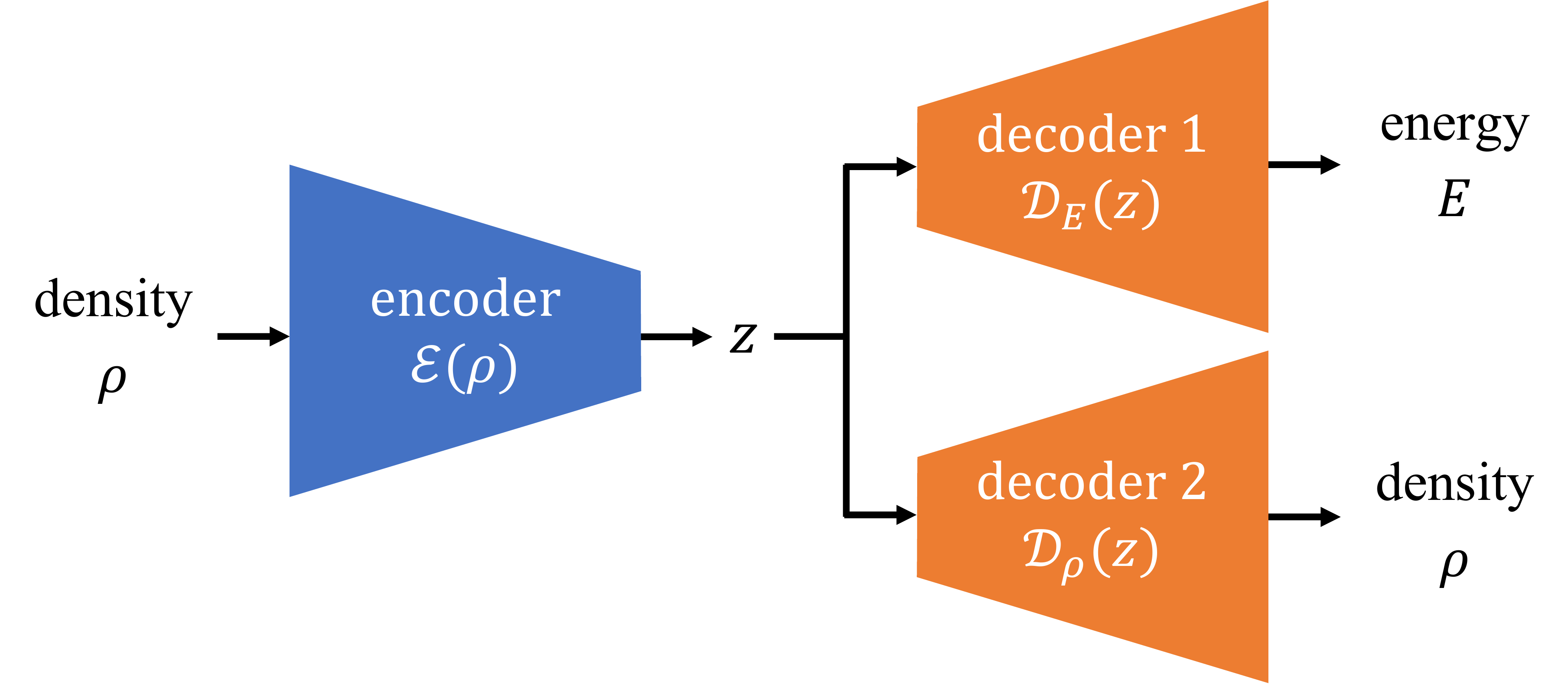}
    \caption{\label{fig:concept}
    A conceptual diagram of our model.
    The encoder $\mathcal{E}$ compresses the information on the density $\rho$ into the latent variables $z$, 
    and the decoder of the density  $\mathcal{D}_{\rho}$  reconstructs the density itself from the latent variables.
    At the same time the other decoder $\mathcal{D}_{E}$ predicts the energy from the latent variables.
    }
    \end{center}
\end{figure}

To avoid such drawbacks, we combine these two tasks:
\begin{equation}
    \rho\to\bm{z} = \mathcal{E}[\rho]\to E = \mathcal{D}_E(\bm{z})~{\rm and}~\rho = \mathcal{D}_{\rho}(\bm{z}),
\end{equation}
where $\mathcal{D}_{E}$ and $\mathcal{D}_{\rho}$ are decoders for the energy and for the density, respectively. 
The encoder-decoder functions are approximated by neural networks.
Figure \ref{fig:concept} illustrates the conceptual architecture of our model.
We can expect that the common feature representation $\bm{z}$ contains information on both the densities and the energies.

Machine learning is referred to as multi-task learning (MTL) 
when multiple tasks are solved simultaneously \cite{MTL_review}.
While MTL is often utilized to improve the generalization performance of all tasks, 
the primary purpose of our study is to apply MTL to obtain a common feature representation across the tasks.

\subsubsection{architecture}
To perform MTL, one must define a specific model for the encoder and the decoders, $\mathcal{E}$, $\mathcal{D}_{E}$, and $\mathcal{D}_{\rho}$. 
In this work, the input for the neural network is a nucleon density $\rho(r, z)$, 
that is, the binding energy is determined only by the total density, $\rho = \rho_p + \rho_n$, 
rather than the proton and neutron densities $\rho_p$ and $\rho_n$ separately.  
The total nucleon density $\rho(r, z)$
can be regarded as a monochromatic image 
with the size $1\times48\times128$.
As this size is too large to use in 
an architecture which consists only of 
fully-connected layers, 
we mainly use convolutional neural networks (CNNs), which have been successful in the image recognition field.
We in particular adopt ResNet \cite{ResNet}, which is one of the most famous CNN models and won a 2015 image recognition competition \cite{LSVRC2015}.
The critical idea of ResNet is to use the skip (residual) connection, where an output from a previous layer is added to the output 
of the current layer, allowing for a benefit of multi-layering.
In the original paper of ResNet \cite{ResNet}, five different models with 18, 34, 50, 101, and 152 layers are proposed, which are 
called ResNet18, ResNet34, ResNet50, ResNet101, and ResNet152, respectively.

For the encoder $\mathcal{E}$, we 
adopt the ResNet18 model. 
Since our image data is smaller than the one of ImageNet \cite{ImageNet}, for which the size of images is $3\times 256 \times 256$,  
we remove the 2D max pooling layer with the kernel size = 3, stride = 2, 
and padding = 1 from the original ResNet18. 
The size of each output sample is equal to the latent dimension $d_z$, 
which is a changeable parameter.

\begin{figure*}[tbhp]
    \begin{center}
    \includegraphics[width=172mm]{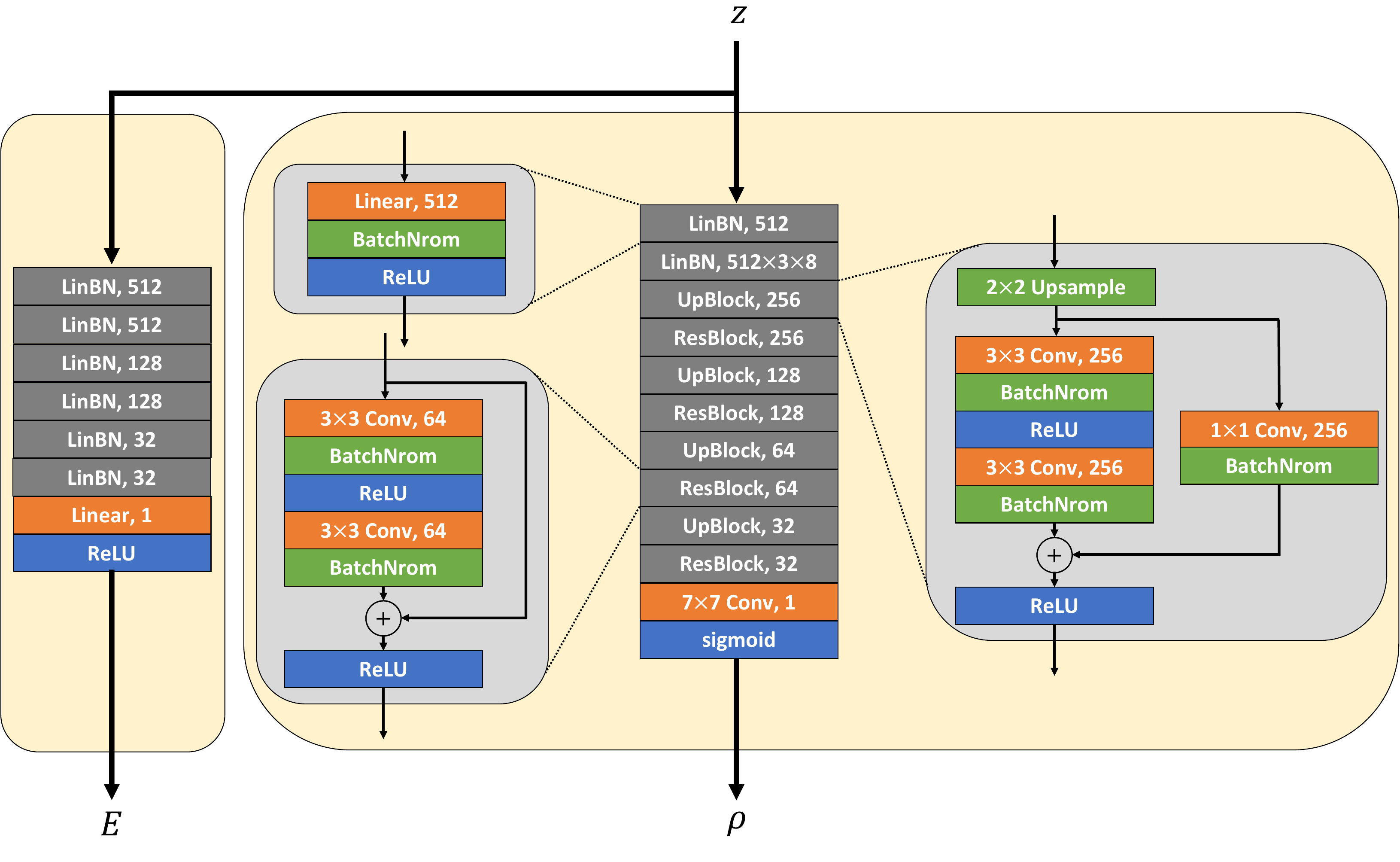}
    \caption{ \label{fig:decoder}
    The architecture of the decoders in our MTL model.
    The left and right blocks correspond to the decoders for the energies, $\mathcal{D}_E$, and for the densities, $\mathcal{D}_{\rho}$, respectively.
    In the convolutional layers, the left numbers denote the kernel sizes, and the right numbers are for 
    the number of the output channels.
    All the amount of padding is taken to be the same padding.
    The upsample layer doubles the size of the image both vertically and horizontally, for which the interpolation is not applied.
    }
    \end{center}
\end{figure*}

For the decoder for the densities, $\mathcal{D}_{\rho}$, we do not use ResNet as it is, but adopt a model motivated by ResNet as shown in Fig. \ref{fig:decoder}\footnote{
We also checked the Parametric Rectified Linear Unit (PReLU) \cite{PReLU} for 
the activation functions, but the performance was not improved.
}.
Note that we do not use the transposed convolution 
because it may cause checkerboard artifacts \cite{checkerboard_artifact}.
On the other hand, upsampling which we employ in our model 
does not create such problem, although one needs 
to pay attention to jaggies.
In addition, the upsample layer does not explicitly make a specific direction along $z$ axis, 
which is physically reasonable for the axial symmetric system studied in this paper.
For the decoder of the energy, $\mathcal{D}_{E}$, 
we use a model with six fully-connected layers (see Fig. \ref{fig:decoder}). 
Notice that we use a relatively small number of layers in this study 
to avoid an overcomplicated dependence of the energy on the latent variables. 

\subsubsection{loss function}
A multi-task learning involves several tasks and an appropriate optimization is a critical problem.
A common approach is to balance the individual loss functions for separate tasks \cite{MTL_review}. 
Several approaches have been proposed to combine these different loss functions, such as uncertainty weighting (UW) \cite{UW} and dynamic weight average (DWA) \cite{DWA}.
In our MTL model, the energy prediction task and the density prediction task tend to compete with 
each other, especially when the latent dimension $d_{z}$ is small.
It would be desirable to define a single metric to evaluate the performances of these tasks, but 
it is difficult to determine it in advance. 
Therefore, in this paper we simply combine an energy loss $\mathcal{L}_{E}$ 
and a density loss $\mathcal{L}_{\rho}$ with 
a constant ratio $\alpha$ as 
\begin{equation}
    \label{eq:MTL_loss}
    \mathcal{L}_{\rm{MTL}}(D;\alpha)
    =
    \mathcal{L}_{E}(D) +
    \alpha\,\mathcal{L}_{\rho}(D), 
\end{equation}
 with    
\begin{equation} 
    \label{eq:E_error}
    \mathcal{L}_{E}(D)
    =
    \frac{1}{|D|}\sum_{i=1}^{|D|}\left(
        E^{(i)}_{\rm{pred}} - E^{(i)}_{\rm{true}}
    \right)^2,
\end{equation}
and
\begin{equation}
    \label{eq:rho_error}
    \mathcal{L}_{\rho}(D)
    =
    \frac{1}{|D|}\sum_{i=1}^{|D|}
    \int d^3r\left(
        \rho^{(i)}_{\rm{pred}}(\bm{r}) - \rho^{(i)}_{\rm{true}}(\bm{r})
    \right)^2,
\end{equation}
where $|D|$ is the number of data in a dataset $D$.
The constant hyperparameter $\alpha$ determines the relative importance 
between the energy and the density for optimization. We will vary this parameter to examine 
the performance of our MTL as we will show in the next section. 
All the parameters in the neural networks are optimized to minimize the loss function $\mathcal{L}_{\rm{MTL}}$.
Note that in the actual calculations, the data are min-max normalized, even though 
we return to the original physical normalization when we discuss a performance of MTL. 
Then the energies $E$ and the densities $\rho$ are also normalized, 
and the loss function for the energies $\mathcal{L}_{E}(D)$ is dimensionless, but 
that for the densities $\mathcal{L}_{\rho}(D)$ has a physical dimension $\mathrm{fm}^3$ originated 
from the radial integral in Eq. (\ref{eq:rho_error}). 
Therefore, the physical dimension of $\alpha$ is $\mathrm{fm}^{-3}$.

\subsubsection{optimization}
To optimize the parameters in the neural networks, 
we utilize Adam optimizer \cite{Adam} with the default hyperparameters 
in Pytorch \cite{Pytorch}, except for the learning rate.
The initial value of the learning rate is $10^{-4}$ and, 
following Ref. \cite{VS22}, we dynamically reduce the learning rate by 
a factor of 0.5 if the loss of validation data is not improved among the 
last 15 epochs. 
One can easily implement this with "torch.optim.lr\_scheduler.ReduceLROnPlateau".
When using such approach, that is, 
the approach with a dynamic learning rate decay, 
it is not fair to evaluate the final performance 
with the validation data as the approach itself depends upon the data.
Therefore, we randomly divide our dataset into training data (90\%), validation data (5\%), and test data (5\%), among which the test data are utilized only for a final evaluation of the performance.

In our model, 
the mini-batch size is 128 and the number of epochs is 1000.
Since the convergence of the validation loss is worse than that of the training loss in our MTL calculations, 
and the validation loss sometimes fluctuates during learning, 
we adopt the parameters which yield the smallest loss $\mathcal{L}_{\rm{MTL}}$ for the validation data 
and use them to evaluate the results. 
As we will show in the next section, we will also perform other type of deep learning calculations in this 
paper. Even in that case, 
we use the same optimization method unless otherwise mentioned.  

It is useful to mention the numerical cost of our calculations: 
optimizing our MTL model for a single set of hyperparameters takes 
50 GPU-hours with NVIDIA RTX 3090TI.
Namely, we must run $n$ NVIDIA RTX 3090TIs for $50/n$ hours.

\section{Criteria for evaluation}
\label{sec:criteria}

\subsection{Criterion for energy}
\label{sec:criterion-energy}

Before we evaluate results of MTL, let us first discuss criteria for such evaluations. 
We first discuss errors for the energies. 
Because the number of our data is finite, it is natural that the energy cannot be perfectly predicted,
and thus such training should not be done in terms of generalization performance.
Therefore, we first estimate how much errors our dataset admits in terms of energy prediction.
Namely, we learn the OF-EDF $E[\rho]$ using supervised learning, and regard the resultant error as a limitation 
of the energy prediction.

Table \ref{tab:criterion} summarizes the mean absolute errors (MAEs) and the 
mean square errors (MSEs) for the energy prediction with several neural networks.
These are defined as 
\begin{equation}
    \mathrm{MAE}
    =
    \frac{1}{|D|}\sum_{i=1}^{|D|}
    \left|
        E^{(i)}_{\rm{pred}} - E^{(i)}_{\rm{true}}
    \right|, 
    \label{eq:MAE}
\end{equation}
and 
\begin{equation}
    \mathrm{MSE}
    =
    \frac{1}{|D|}\sum_{i=1}^{|D|}
    \left|
        E^{(i)}_{\rm{pred}} - E^{(i)}_{\rm{true}}
    \right|^2, 
\end{equation}
respectively. 
\begin{table*}[tbhp]
    \caption{
    The mean absolute error (MAEs) and the 
    mean square error (MSEs) for supervised learning  of OF-EDF $E[\rho]$ with ResNets and ViT. 
    The results imply that 64 keV is a criterion for energy predictions.
    For comparison, we also perform another training with proton and neutron number densities as inputs,  
    as shown in the last row. 
    }
    \label{tab:criterion}
    \centering
    \begin{tabular}{|c|c|c|r|r|r|r|} \hline
        &  &  & \multicolumn{2}{|c|}{validation data (5\%)} & \multicolumn{2}{|c|}{test data (5\%)} \\ \cline{4-7}
        model & input(s) & output & MAE / MeV & MSE / MeV$^2$ & MAE / MeV & MSE / MeV$^2$ \\
       \hline
        ResNet18 & $\rho$ & $E$ &0.0633 & 0.00965 & 0.0638 & 0.00963\\
        ResNet34 & $\rho$ & $E$ & 0.0638 & 0.00965 & 0.0646 & 0.00949 \\
        ResNet50 & $\rho$ & $E$ & 0.0649 & 0.01084 & 0.0649 & 0.01023\\
        ResNet09 & $\rho$ & $E$ & 0.0724 & 0.01197 & 0.0720 & 0.01161\\
        ViT \cite{ViT} & $\rho$ & $E$ & 0.2613 & 0.15161 & 0.2704 & 0.16302 \\
        ResNet18 & $\rho_n$, $\rho_p$ & $E$ & 0.0641 &0.00973 &0.0635 &0.00951\\
        \hline
    \end{tabular}
\end{table*}
ResNets are the same as those described in Sec. II, except for the output layer, for which 
we operate ReLU at the last stage. 
ResNet09, which is not included in the original paper \cite{ResNet},  
is a model in which the residual blocks are halved from ResNet18.
We optimize each of the models in Table \ref{tab:criterion} using the MSE loss function, Eq. (\ref{eq:E_error}).
We notice that the errors of ResNet34 and ResNet50 are almost that same as that of ResNet18, 
while the error of ResNet09 is significantly worse, implying that ResNet18 has an enough 
representational capacity to capture the characteristics of our dataset.
Because of this, 
we mainly use ResNet18 in this paper. 
We thus define the criterion for the energy prediction as 64 keV.

We also evaluate the OF-EDF with Vision Transformer (ViT) \cite{ViT}, 
which has achieved the state-of-the-art performance in the image recognition. 
ViT is an application of the transformer \cite{Transformer}, which has been remarkably successful, especially in large language models (LLMs), to image recognition. 
We utilize a ViT model consisting of 16 transformer blocks with multi-head self attention in which 
the number of heads and the dimension of hidden layers are 16 and 2048, respectively.
The patch size, the embedding dimension, and the dropout rate are $8\times 8$, 128, and 0.3, respectively.
Table \ref{tab:criterion} indicates that 
the errors of our ViT model are much larger than those with the ResNets.  
However, it is important to point out that ViT requires a large amount of data to outperform CNN models \cite{ViT}. 
Once 
a large amount of data suitable for machine learning will become available in nuclear physics, 
we expect that 
ViT will play an important role. Therefore, it is meaningful to study 
the performance of ViT at this stage. 

The calculations presented so far are performed with the total densities as inputs. 
To validate such calculations, 
we also perform another training with the proton and the 
neutron number densities as inputs. 
The result is shown in the last row in Table \ref{tab:criterion}, which indicates 
that the errors are similar to those with the total densities. 

\subsection{Criterion for density}

\begin{figure}[tbp]
    \begin{center}
    \includegraphics[width=86mm]{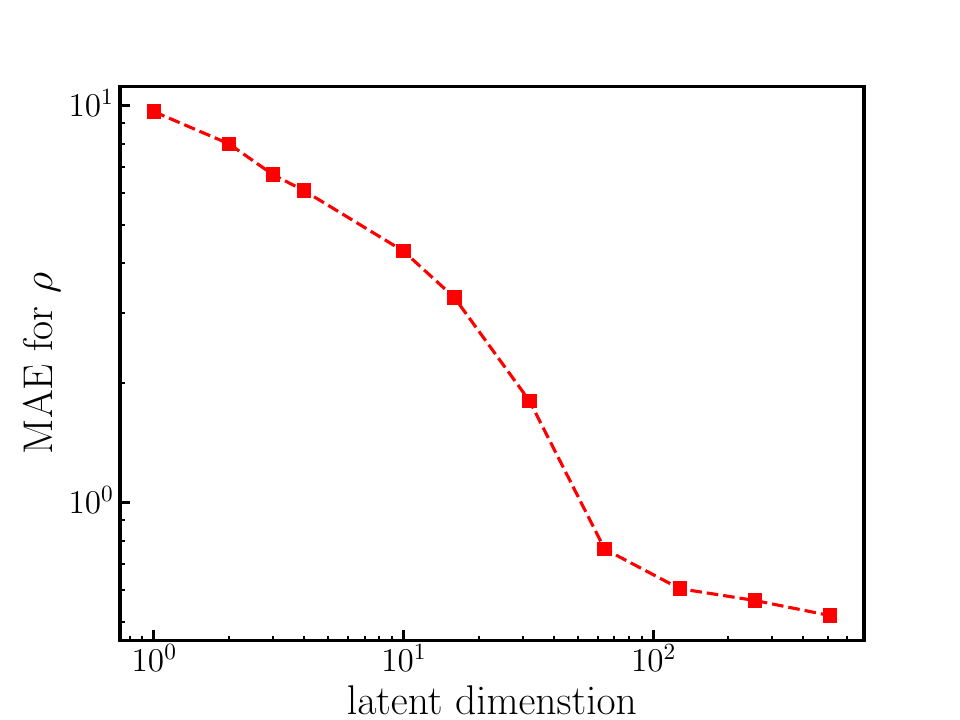}
    \caption{\label{fig:loss_rho}
    The mean absolute errors (MAEs) for a predicting density with the autoencoder 
    as a function of the latent dimension $d_z$, taken as $d_z=1, 2, 3, 4, 10, 2^4, 2^5,2^6, 2^7, 2^8, $ and $2^9=512$.
    }
    \end{center}
\end{figure}

Let us next evaluate the errors in the densities.
It is important to notice that density is an input variable in the DFT, and a 
supervised learning cannot be used to evaluate the errors.
Therefore, we use the autoencoder (AE) \cite{recnet_AE}, which is a kind of unsupervised learning.
In our MTL shown in Fig. \ref{fig:concept}, if we remove the decoder for 
the energy, $\mathcal{D}_E$, as well as 
the loss function for the energy, the model is equivalent to the AE for the 
density (see Eq. (\ref{eq:learning-rho})).
In this way, we evaluate the AE for several latent dimensions. 
The results are shown in Fig. \ref{fig:loss_rho}.
Here, the MAE for the density is defined as
\begin{equation}
    \mathrm{MAE}
    =
    \frac{1}{|D|}\sum_{i=1}^{|D|}
    \int d^3r\left|
        \rho^{(i)}_{\rm{pred}}(\bm{r}) - \rho^{(i)}_{\rm{true}}(\bm{r})
    \right|.
\end{equation}
Note that the MAE is a dimensionless variable as the densities in this equation has physical dimension. 
One can see that the MAEs in lower dimensions are as large as a few percent of the mass number, $A=236$.
These large errors imply that our dataset contains a variety of densities. 
Because of this, increasing the latent dimension does not rapidly reduce the errors, even though 
the errors are significant small when the latent dimension is large: 
MAE of 0.5 is achieved for $d_z = 512$, that is the maximum dimension for the original ResNets \cite{ResNet}.  
For each $d_z$, we will use the corresponding MAE for a criterion for density. 

\subsection{Errors with multipole moments}

\begin{figure}[tbp]
    \begin{center}
    \includegraphics[width=86mm]{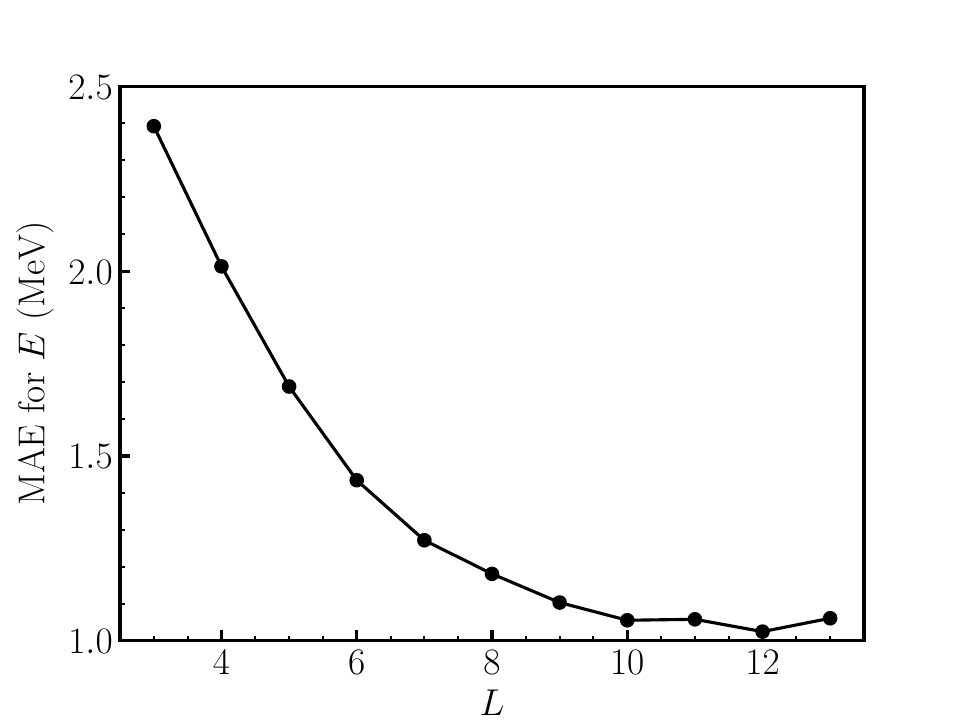}
    \caption{\label{fig:Q_error}
    The mean absolute errors (MAEs) for predicting energy from multipole moments.
    The vertical axis denotes the maximum multipoles in $S_{L} =\{Q_{20}, Q_{30}, \cdots, Q_{L0}\}$, 
which is similar to the latent dimension 
    in our MTL. 
    }
    \end{center}
\end{figure}

Conventionally, the multipole moments $\{Q_{\lambda, k}\}$ have often been 
employed to describe nuclear shape dynamics.  
It is therefore interesting to evaluate with our dataset 
how much information  
these variables contain. 
To this end, we apply a supervised learning to predict the energies from several sets of multipole moments,  $E(Q_{20}, Q_{30}, \cdots, Q_{L0})$.
A neural network which we use is the decoder of our MTL model, $\mathcal{D}_{E}$, 
with $\{Q_{\lambda, k}\}$, instead of the latent variables $z$, as inputs. 
Figure \ref{fig:Q_error} shows the MAEs for several sets of the input variable 
defined as $S_{L} =\{Q_{20}, Q_{30}, \cdots, Q_{L0}\}$.
One can see that these errors are significantly larger than the criterion defined 
in Sec. \ref{sec:criterion-energy}, that is, 64 keV.
We have also tried with a deeper model with more trainable parameters, 
but the results were not improved.
As a reference, we compute the MAE by assuming 
that the output energies are given by uniform random numbers within a range of energies in our dataset, 
that is, the MAE obtained by replacing 
$E^{(i)}_{\rm{pred}}$ in Eq. (\ref{eq:MAE}) with random numbers. 
The resultant value is 5.0 MeV. 
The errors with the low order multipole moments are reduced from this value only by a factor of about 2, 
indicating that the multipole moments do not contain enough information, at least for our dataset. 

We also examine the density prediction from the multipole moments using 
the decoder $\mathcal{D}_{\rho}$. 
We find that 
the mean absolute errors are 
several times larger than the mass number $A = 236$, 
indicating that the multipole moments do not have a capacity to predict the densities, 
at least, for our dataset.
Our results might be somewhat improved with techniques such as 
pre-training with other dataset, though. 
Rather than attempting it, however, in the next section we will propose 
to use MTL to extract much more consistent variables for shape dynamics than the multipole 
moments. 

\section{Results of MTL}

\subsection{Latent variables and dimensions}

\begin{figure*}[thbp]
    \begin{center}
    \includegraphics[width=172mm]{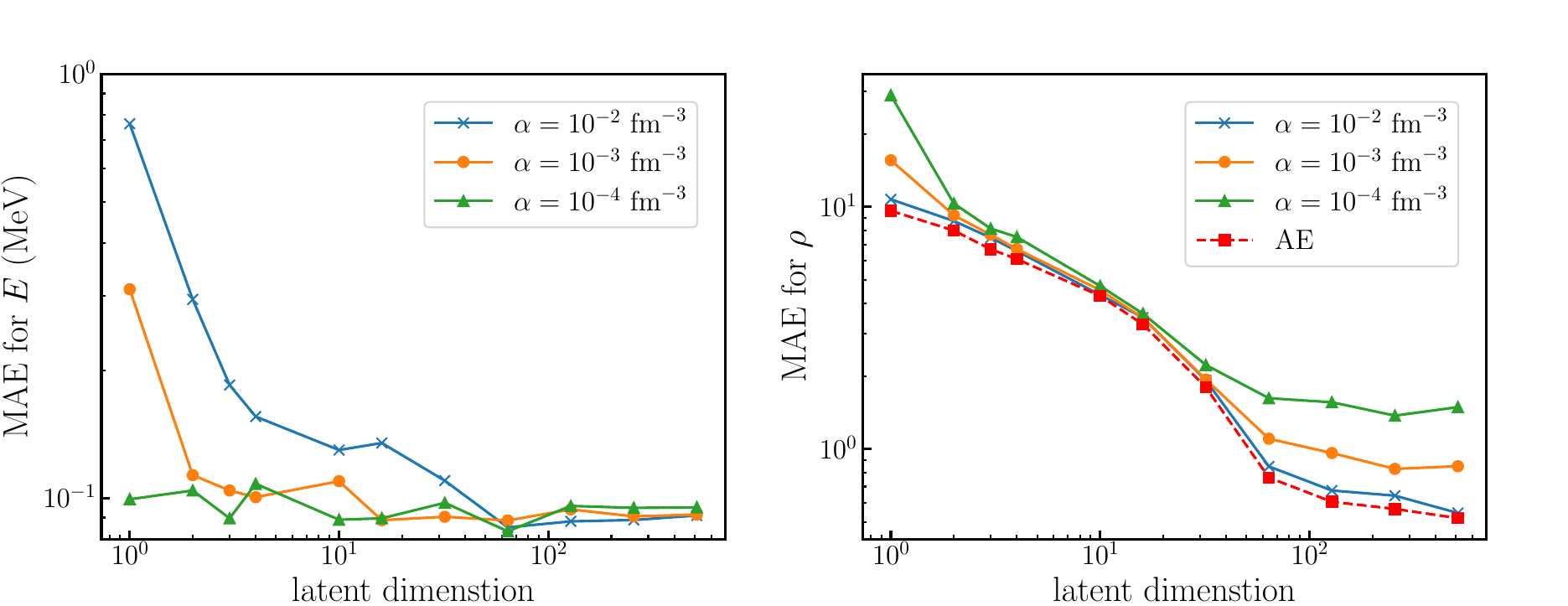}
    \caption{\label{fig:MTL_E}
    The mean absolute errors (MAEs) of the densities and the energies as a function of the latent dimension for several  values of the hyperparameter $\alpha$ in the MTL loss function given by Eq.(\ref{eq:MTL_loss}). 
    The dashed line on the right panel is the same as 
    that plotted in Fig. \ref{fig:loss_rho}, which shows the results of the autoencoders (AE).
    }
    \end{center}
\end{figure*}

\begin{figure*}[tbhp]
    \begin{center}
    \includegraphics[width=170mm]{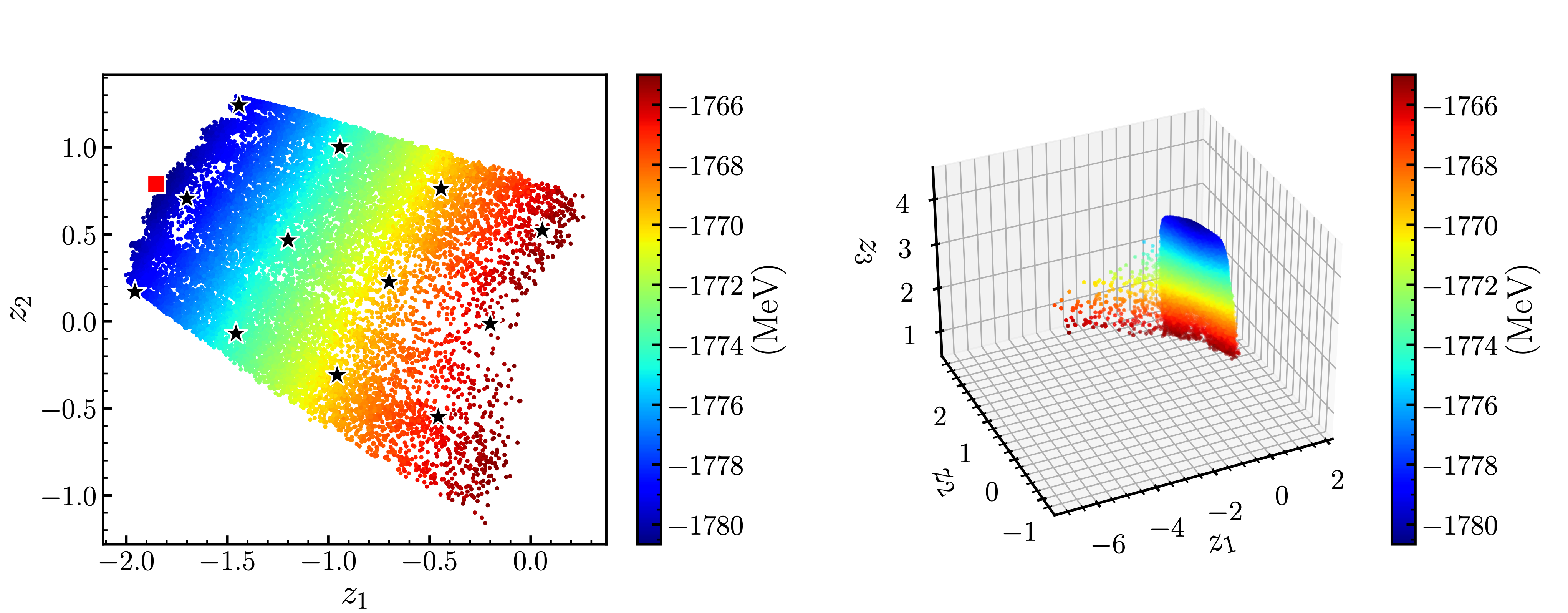}
    \caption{\label{fig:latent}
    The latent variables for a combination of the validation data (5\%) and the test data (5\%).
    The performance for these data is almost the same to each other, and 
    we include the validation data in this figure only for a presentation purpose. 
    The latent dimension is 2 (the left panel) and 3 (the right panel), 
    and the hyperparameter of the MTL loss function is set to be $\alpha=10^{-3}$  $\mathrm{fm}^{-3}$. The binding energy of each data are denoted with color.
    In the left panel, the ground state with no external field is shown by a filled square, for which the corresponding latent variables are $(z_1,z_2)=(-1.85,  0.79)$.
    The black stars denote those points whose densities are 
    reconstructed from the latent variables in Fig. \ref{fig:sample}. 
    }
    \end{center}
\end{figure*}

Let us now discuss the MTL results and the extracted latent variables.
The MTL loss function has one hyperparameter $\alpha$, 
which determines the relative priority between 
the errors of energy and those of density (see Eq. (\ref{eq:MTL_loss})).
We determine the value of $\alpha$ so that $\mathcal{L}_E$ and $\mathcal{L}_{\rho}$ 
are roughly on the same scale.
In this study, we employ three different values for $\alpha$, that is, 
$\alpha=10^{-2}$, $10^{-3}$, 
and $10^{-4}$  $\mathrm{fm}^{-3}$. 

Figure \ref{fig:MTL_E} illustrates the MAE for the energies (the left panel) and for the densities (the right panel).
One can see that increasing the latent dimension rapidly increases the performance of the energy prediction, 
while the performance for the densities tends to increase only slowly.
As the two competing tasks tend to increase the errors 
of each other, the density and energy errors behave oppositely 
when increasing or decreasing $\alpha$.
That is, while increasing $\alpha$ reduces the errors in the densities, it increases the errors in the energies.

Remarkably, even with a small number of the latent variables, the energy can be predicted within errors of at most a few hundred keV.
In particular, when the latent dimension is increased from 1 to 2, 
the errors in energy decrease rapidly for $\alpha=10^{-2}$ and 10$^{-3}$ fm$^{-3}$. 
With two latent variables, the densities can also be predicted well: the density prediction errors 
for $d_z=2$ are in fact larger than the results of 
the autoencoders (AE) only by a small amount (see the dashed line).

It is a good compromise to set $\alpha=10^{-3}$ fm$^{-3}$. For larger values of $\alpha$, 
the density prediction is better but the energy prediction is worse. 
On the other hand, for small values of $\alpha$ the density predication is worse. 
With $\alpha=10^{-3}$ fm$^{-3}$, both the energy predication and the density prediction 
are simultaneously good. 
With this choice of $\alpha$,
the errors in energy do not decrease much when the latent dimension is 
increased from 2 to 3. 
This implies 
that the main part of the dynamics in our dataset can be characterized only by two parameters.
We emphasize that 
our latent variables contain much more information on our dataset 
as 
compared to the conventional multipole moments.  
This can be regard as an evidence that the multipole moments 
may not provide appropriate collective coordinates to describe shape dynamics of heavy nuclei, such as 
$^{236}$U, even in the region of small deviations from the ground state deformation. 

\begin{figure*}[tbhp]
    \begin{center}
    \includegraphics[width=170mm]{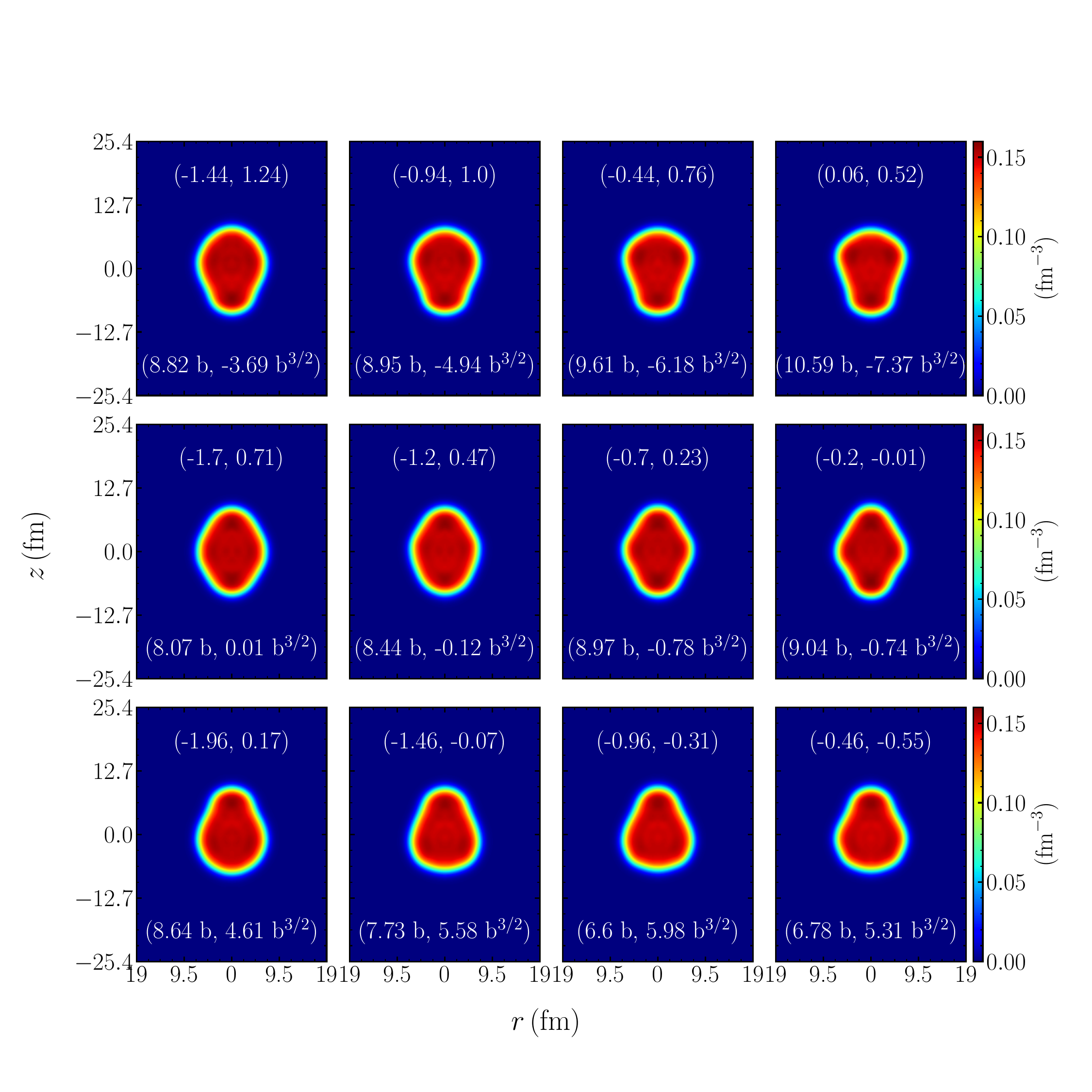}
    \caption{\label{fig:sample}
    Sample densities reconstructed from latent variables for the points shown in Fig. \ref{fig:latent}. 
    The two numbers displayed on the top of each panel denote the latent variables ($z_1$,  $z_2$), while those at the bottom are the quadrupole and the octupole moments, $(Q_{20}, Q_{30})$.
    }
    \end{center}
\end{figure*}

In the cases where the latent dimension is smaller than 3, 
we can easily visualize the profile of the latent variables.
Especially, the 2 and 3 dimensional cases have a good performance 
compared with the criteria defined in Sec. \ref{sec:criteria}. 
Figure \ref{fig:latent} shows scatter plots of the latent variables obtained 
with $d_z=2$ (the left panel) and $d_z=3$ (the right panel) together with 
$\alpha=10^{-3}$  $\mathrm{fm}^{-3}$. 
Notice that the latent variables do not have a physical dimension, 
and the scales in the figure are irrelevant. 
The color dimension denotes the binding energy of each data. 
Here, we plot both the validation data and the test data, since the errors for those data 
are almost the same.
The ground state of $^{236}$U is at $(z_1,z_2)=(-1.85,  0.79)$ if the external potential 
is not introduced, that is shown by a square dot in the left panel. 
One can notice that there is a symmetric axis in the plots.
To see the significance of the symmetric axis, the reconstructed densities are plotted 
in Fig. \ref{fig:sample} for the points shown in the left panel of 
Fig.  \ref{fig:latent}.  
Even though those densities contain certain errors as they are reconstructed ones, 
one can still see that 
the shapes on the opposite side across the symmetry axis 
can be regarded as the same shapes but inverted with respect to the $z$ axis.
This implies that our MTL model successfully recognizes the parity symmetry.
Note that the direction of the symmetry axis and the position of the origin 
are not physically meaningful, as they are determined by the initial trainable parameters of the neural network.
In fact, we have performed exactly the same calculations but with another initial condition, and obtained 
a similar scatter plot with a different position of the origin of the $z$ coordinate.

\begin{figure*}[tbhp]
    \begin{center}
    \includegraphics[width=170mm]{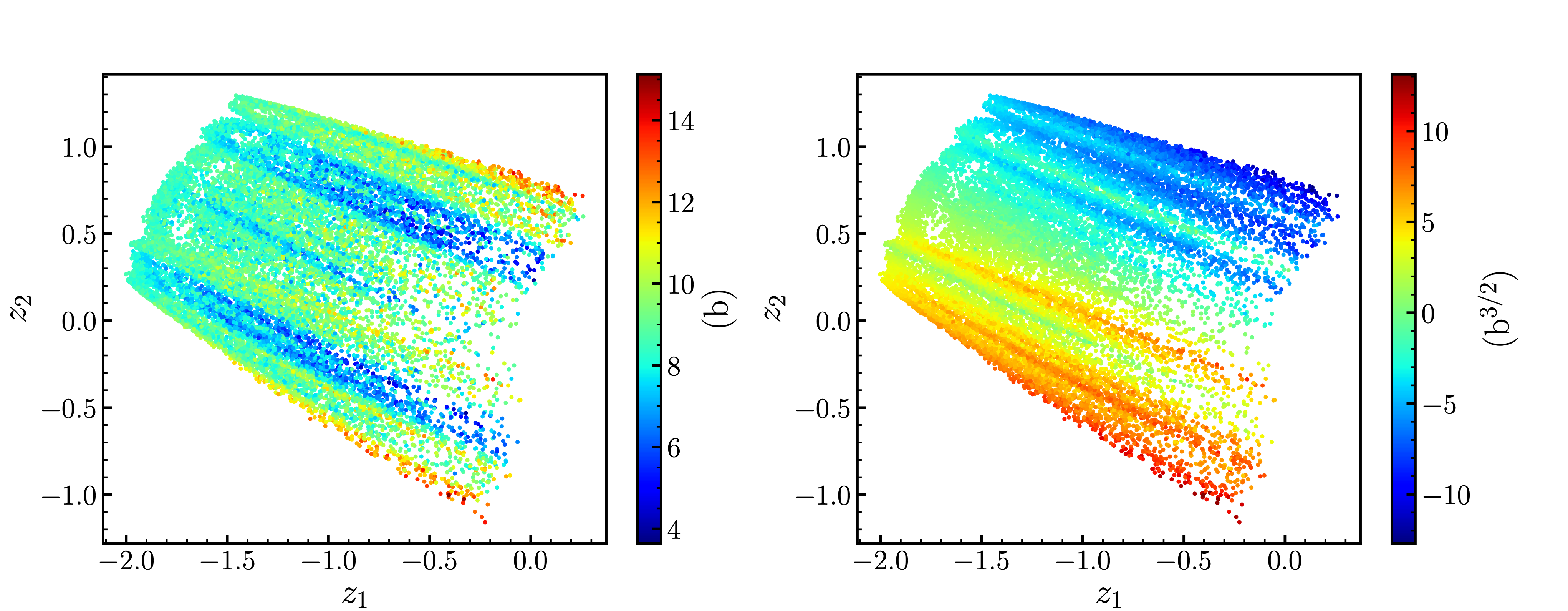}
    \caption{\label{fig:scatter_Q}
    Same as the left panel of Fig. \ref{fig:latent}, but for 
    the quadrupole moment (the left panel) and the octupole moment (the right panel). 
    }
    \end{center}
\end{figure*}

Let us focus on the case with $d_z=2$. 
Figure \ref{fig:scatter_Q} shows the quadruple moment and the octupole moment of each latent variable. 
One can see the existence of the symmetry axis.
The figure also indicates that $Q_{20}$ and $Q_{30}$ 
do not play a special role in our dataset.
An advantage of our latent variables is 
that all the points which correspond to excited states for given multipole moments 
can be treated as ground states at 
the corresponding latent variables. 
Even though it is trivial to obtain a smooth energy surface if the dimension of the latent variables 
is large, it is notable that this is achieved even with the dimension of 2.  
On the other hand, if the energies are plotted in the $(Q_{20},Q_{30})$ plane, those energies are 
rather scattered as one can see in Fig. Figure \ref{fig:scatter_data}. 
Note that 
all the configurations correspond to the ground state with external fields, but they do not necessarily correspond 
to the local ground state for each $(Q_{20},Q_{30})$ in the constrained HFB.

\begin{figure*}[tbhp]
    \begin{center}
    \includegraphics[width=170mm]{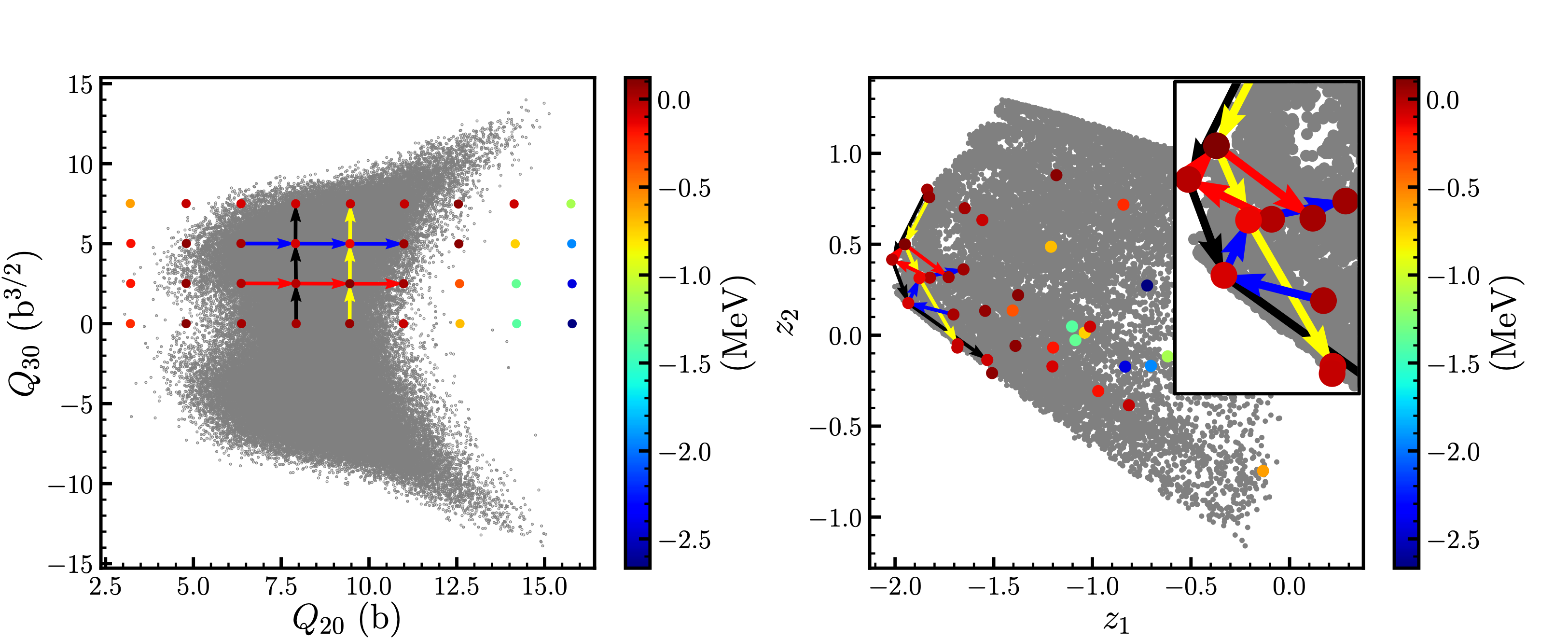}
    \caption{\label{fig:domain_shift}
    Mapping from Fig. \ref{fig:scatter_data} to the left panel of Fig. \ref{fig:latent}. 
    Points connecting with a color arrow on the left panel correspond to the points connecting 
    with the same color in the right panel. The color for each points denotes the difference 
    between the true energy and the predicted energy estimated with the MTL model with $\alpha=10^{-3}$ $\mathrm{fm}^{-3}$ and $d_z = 2$.
    }
    \end{center}
\end{figure*}

To see 
how the multipole moments $(Q_{20},Q_{30})$ 
in Fig. \ref{fig:scatter_data} map onto 
the two dimensional plot for the latent variables shown in the left panel of 
Fig. \ref{fig:latent}, 
we carry out the constrained HFB calculations. 
Points connecting with color arrows on the left panel of Fig. \ref{fig:domain_shift} 
correspond to the points connecting with the same color on the right panel. 
One can see that the trajectories on the $(z_1,z_2)$ plane are quite complex, and the 
mapping from $(Q_{20},Q_{30})$ to $(z_1,z_2)$ is highly non-trivial. 
Evidently, the main feature of our dataset captured by MTL cannot easily be 
described by conventional quadrupole moments and requires 
novel latent variables. 

\subsection{Domain shift}

The fact that a neural network performs well on test data does not necessarily 
mean that the model will perform well when users try with their own data.
This is because properties of two datasets may be different.
To clarify this problem, we introduce the idea of a domain in the context 
of Transfer Learning (TL).
Here, a domain $\mathcal{D} = \{\mathcal{X}, P(X)\}$ consists of a feature (input) space $\mathcal{X}$ and a probability 
distribution $P(X), X\in \mathcal{X}$ \cite{domain_adaptation_survey1, domain_adaptation_survey2} 
\footnote{
Introducing the idea of a task allows 
a more detailed and strict categorization of a problem.
See Refs. \cite{domain_adaptation_survey1, domain_adaptation_survey2} for details. 
}.
In this study, $\mathcal{X}$ corresponds to a set of densities as a whole, 
and $P(X)$ corresponds to the probability distribution of the densities $p[\rho]$ 
discussed in Sec. II-A.
In machine learning, one usually prepares a labeled dataset for training and test 
by sampling from a source domain $\mathcal{D}^s$ and a target domain $\mathcal{D}^t$, respectively.
In traditional machine learning methods, which we follow in this paper, $\mathcal{D}^s$ is assumed to be identical to $\mathcal{D}^t$.
We have generated both the datasets by randomly splitting a single dataset, and 
therefore we have also adopted this assumption.
However, the traditional machine learning methods do not guarantee a good performance 
for a target domain when the domains are different, i.e., $\mathcal{D}^s\ne\mathcal{D}^t$.
Such problem of domain shift 
is quite crucial when one would like to adopt a learned model to 
actual applications.
It is not uncommon that a performance of a model immediately declines when the model is 
actually applied, even when the model shows superior performance with data at hand. 

In this subsection, we check a performance of our MTL model when it is applied to data generated by a 
domain which is different from the one of our dataset, namely, the source domain.
In nuclear physics, the constrained HFB method with $Q_{20}$ and $Q_{30}$ is often utilized, 
and we regard that as a target domain.
Of course the corresponding dataset is expected to have some overlap with the dataset 
generated by our random potentials, but the probability would be 
quite small because the number of data is finite.
Thus, we can consider that 
the target domain is different from the source domain in our dataset.

The colors of the points in Fig. \ref{fig:domain_shift} 
show the difference between the true energy and the predicted 
energy with the MTL model.
In the left panel, 
one can see that the performance of our MTL model is high in the region 
where the train data exist in terms of multipole moments, 
indicating that our model is not overly adapted to the source domain.
On the other hand, 
the performance becomes 
significantly worse outside the region. 

For a fission problem, one would also need a calculation for large deformations which are 
outside the region studied in this paper. 
If we want the MTL model to have a prediction ability even for 
such larger deformations, 
we simply need to use larger datasets that include such deformed states. 
Alternatively, one can also use other techniques, such as 
domain adaptation (DA) \cite{domain_adaptation_survey1, domain_adaptation_survey2} and domain generalization (DG) \cite{domain_generalization_survey}.

\subsection{Symmetry and data augmentation}

\begin{table*}[tbp]
 \caption{
     The metrics $M_E$ and $M_\rho$ defined by Eqs. (\ref{eq:M_E}) and (\ref{eq:M_rho}) 
     to evaluate the role of symmetry of the MTL models.
     These are evaluated only for test data (5\%).
     For comparison, the MAEs for energy and density are also shown. 
     If symmetry is "False", the results for the MAEs are the same as those in Fig. \ref{fig:MTL_E}.
     On the other hand, in the case of "True", we apply the data augmentation as well as 
     the improvement of the encoder.
 }
 \label{tab:symmetry}
 \centering
    \begin{tabular}{|c|c|r|r|r|r|} \hline
    symmetry & $d_z$ & $M_{E}$ / MeV & $M_{\rho}$ & MAE ($E$) / MeV & MAE ($\rho$)  \\
    \hline
    False & 2 & 0.0842 & 5.0618 & 0.1133 & 9.2352 \\
    True & 2 & 0.0798 & 5.4921 & 0.1250 & 9.2756 \\
    \hline
    False & 3 & 0.0675 & 3.7661 & 0.1043 & 7.6681 \\
    True & 3 & 0.0716 & 3.3035 & 0.1003 & 7.7055 \\
    \hline
    \end{tabular}
\end{table*}

Even though the latent variables obtained in this study are good enough, 
one may obtain better latent variables by taking into account 
symmetry of the datasets.
To check this, in this subsection we repeat MTL by constructing the 
datasets in which symmetry of the system is respected.

The densities constructed in this paper have 
axial symmetry along the $z$ axis. Even though those densities are in general 
break parity symmetry, the binding energy is invariant with respect to the parity operation.  
To reflect this physical symmetry in the learning results, 
it is desirable to have the same symmetry in the dataset as well.
Since the source domain is invariant under the parity transformation, 
if one samples a quite large amount of data, 
one can assume that 
the distribution of the dataset should also be invariant under the parity transformation.
However, this is difficult in practice with finite amount of data. 
Therefore, we adopt the data augmentation of flipping image, for which the corresponding 
energy is exactly invariant due to the parity symmetry.
In the actual calculations, we randomly flip a input image with respect to 
$z$ axis with a probability of 0.5.
In addition, we also introduce an improvement to the encoder of our MTL model. 
That is, 
as the original ResNets, which utilize odd kernel sizes 
even with a stride of 2, 
might cause a shift of a center of the image, 
we increase the kernel sizes by 1 for the convolutional layers 
when the stride is set to be 2.

We train the improved MTL model with the augmented dataset, 
with $d_z=2, 3$ and $\alpha=10^{-3}$ $\mathrm{fm}^{-3}$.
We evaluate the performance of the symmetry with the following two metrics:
\begin{gather}
    M_{E}
    = \sum_{i=1}^{|D|}
    \frac{1}{|D|}
    \left|\mathcal{D}_E(
        \mathcal{E}[\rho^{(i)}]
        )
    - 
    \mathcal{D}_E(
        \mathcal{E}[\rho_\mathcal{F}^{(i)}]        
        ) \right|, \label{eq:M_E}  \\
    M_{\rho}
    = \sum_{i=1}^{|D|}
    \frac{1}{|D|}
    \int d^3r
    \left|\mathcal{D}_\rho(
        \mathcal{E}[\rho^{(i)}]
        )
    - 
    \mathcal{F}[\mathcal{D}_{\rho}(
        \mathcal{E}[\rho_\mathcal{F}^{(i)}]       
        )]
    \right|, \label{eq:M_rho}
\end{gather}
where $\mathcal{F}$ is the flipping operation, 
and $\rho_\mathcal{F}$ is defined as $\mathcal{F}[\rho]$. 
If the model fully respected the symmetry, both of these metrics 
would be exactly zero.

Table \ref{tab:symmetry} shows $M_E$ and $M_{\rho}$ for our datasets. 
One can see that the symmetry consideration does not significantly 
affect the performance of MTL.
This is because our MTL models have enough representation ability, and the datasets are from the beginning 
symmetric enough to learn the symmetry. 
In fact, comparing the two metrics to the MAEs, 
the symmetry is correctly learned within the errors because the metrics are smaller.
Rather, the influence of the initial values of the neural network is more significant.
On the other hand, when we apply the data augmentation to supervised learning 
for the energy only with the improved ResNet18, we find that MAE is reduced to $52$ keV.
Therefore, we conclude that, while data augmentation itself is helpful in increasing accuracy, 
it does not necessarily contribute to the accuracy of models when they 
include adversarial tasks, such as our multi-task learning.

\section{Summary and future perspective}
We have applied multi-task learning (MTL) to a shape dynamics in the vicinity of the ground state of ${}^{236}$U in order to extract a common feature representation of densities and binding energies. 
To this end, we have employed 
the Skyrme Kohn-Sham Density Functional Theory (Skyrme KS-DFT) 
with random external potentials.
In our MTL models, the input densities are compressed into latent variables by an encoder, 
and they are reconverted into energies and densities by decoders.
The resultant latent variables can be regarded as a kind of collective coordinates, 
as the manifold hypothesis and an assumption of collective coordinates are conceptually close 
to each other. 
We have shown that 
latent variables with dimension of as small as 2 
well reproduce the energies and the densities of the test data. 
On the other hand, we have shown that conventional multipole moments contain 
much less information on shape dynamics in our dataset.
This suggests that it is important to choose collective coordinates 
by appropriately taking into account dynamics, like in our latent variables.

In this study we have utilized the autoencoder only for an analysis of latent variables.
When one would like to apply our idea as a generative model, one should adopt 
probabilistic models, such as a variational autoencoder (VAE) \cite{VAE, recnet_AE}.
Notice that, in an image generation, it is difficult to obtain generalization performance 
and generate novel images which are not included in a training dataset.
To handle this, 
Generative Adversarial Networks (GANs) \cite{GAN, GAN_review} and, in the latest studies, Diffusion Models (DMs) \cite{DM_original, DDPM, diffusion_model_survey} are gaining popularity in this field.
In particular, the DMs have achieved the state of the art performance 
for a text to image generation tasks.
Therefore, one may think of applying the latest probabilistic models to nuclear physics.

To extract latent variables that consistently describe nuclear fission, 
it would be necessary to collect a large number of data on larger deformed states.
In that case, the dimension of latent variables may increase, 
even though the dimension of 2 is sufficient for shape dynamics near the ground state. 
However, we found it difficult to obtain such data with random potentials, 
especially data at unstable locations such as a barrier top.  
To obtain such data, one would need either a huge amount of efforts or a drastic improvement of the numerical algorithm 
for convergence of the DFT.
In addition, since the calculations have been performed 
in the framework of density functional theory, 
no information on the many-body wave functions are available from our MTL, and thus 
our method cannot be connected to GCM or other methods that deal with many-body 
wave functions. 
Clearly, novel ideas are necessary 
in order to connect the latent variables to collective models so that e.g. the moment of inertia 
can be computed. 

One of the important conclusions in this paper is that 
sufficiently large data sets themselves contain a wealth of information on the dynamics.
If one analyzes them in an appropriate manner, one can immediately extract information on the 
dynamics. 
Especially in the context of KS-DFT, one can 
avoid using phenomenological constraining fields, 
which are only convenient mathematical tools in most cases in nuclear physics.
This merit enables one to recast nuclear physics from a new perspective.

Fortunately, due to the remarkable growth of the machine learning field in recent years, 
there have been many useful tools to analyze big data.
Unfortunately, however, there is still a lack of datasets to learn in nuclear theory.
Therefore, at this stage, the generation of high quality datasets 
such as ImageNet \cite{ImageNet} is still awaited.
For such data, Vision Transformer (ViT) \cite{ViT} is expected to perform 
better than the conventional neural network models.
In addition, it is also difficult at this moment to train models with more than a billion parameters.
This is because CPUs have been mainly utilized in nuclear physics 
and it may still be difficult to access to a large number 
of expensive GPUs. 
Moreover, it is also important to secure GPUs, which will be even in a greater demand 
in future.

\section*{Acknowledgments}
We thank A. Sannai for useful advise on multi-task learning. We also thank K. Yoshida for discussions. 
A part of numerical computations in this work was carried out at the Yukawa Institute Computer Facility.
This work was supported by JSPS KAKENHI Grants No. JP21KJ1697, No. JP19K03861 and No. JP23K03414.

\bibliographystyle{apsrev4-2}
\bibliography{ref_DL, ref_nucl}

\end{document}